\documentclass[aps,llncs,twocolumn]{revtex4-1} 
\usepackage{hyperref} 
\usepackage[bottom]{footmisc}
\usepackage{graphicx}
\usepackage{amsmath}
\usepackage{amssymb}
\usepackage{float}
\usepackage{subcaption}
\usepackage{braket}
\usepackage{lipsum}
\usepackage{color}
\begin{document}

\title{Field induced hysteretic structural phase switching and possible CDW in Re-doped MoTe$_2$} 

\author{Aastha Vasdev$^1$}
\author{Suman Kamboj$^1$}

\author{ Anshu Sirohi$^1$, Manasi Mandal$^2$, Sourav Marik$^2$, Ravi Prakash Singh$^2$}

\author{Goutam Sheet$^1$}
\email{Corresponding author: goutam@iisermohali.ac.in}

\affiliation{
$^1$Department of Physical Sciences, 
Indian Institute of Science Education and Research Mohali, 
Mohali, Punjab, India}
\affiliation{$^2$Indian Institute of Science Education and Research Bhopal, Bhopal, 462066, India}

\begin{abstract}
\section*{Abstract}
Novel electronic systems displaying exotic physical properties can be derived from complex topological materials through chemical doping. MoTe$_2$, the candidate type-II Weyl semimetal shows dramatically enhanced superconductivity up to 4.1 K upon Re doping in Mo sites. Based on bulk transport and local scanning tunneling microscopy (STM) here we show that Re doping also leads to the emergence of a possible charge density wave (CDW) phase in Re$_{0.2}$Mo$_{0.8}$Te$_2$. In addition, the tunneling $I-V$ characteristics display non-linearity and hysteresis which is commensurate with a hysteresis observed in the change in tip-height ($z$) as a function of applied voltage $V$. The observations indicate an electric field induced hysteretic switching consistent with piezoelectricity and possible ferroelectricity. 
\end{abstract}
\maketitle
\section*{\hspace*{-5cm}{Introduction}}

The transition metal dichalcogenides have attracted renewed attention due to their newly discovered exotic quantum properties. The members belonging to this family have shown diverse topological states of matter including topological insulator \cite{Motohiko,Xiaoyin,Yandong}, Dirac semimetals \cite{Noh,Amit,Shekhar}
and Weyl semimetals\cite{Qi,Liu}. Among the dichalcogenides, MoTe$_2$ has attracted special attention because it exists in diverse structural forms and displays a variety of intriguing physical properties including Weyl physics\cite{Revolinsky,Deng,Tamai,Jiang,Yan} and superconductivity\cite{Qi, Guguchia}.  MoTe$_2$ exists in a trigonal prismatic 2H phase\cite{Puotinen} which is a semiconductor\cite{Albert,Vellinga,Dawson}, the monoclinic 1T$'$ phase which is a semimetal\cite{Brown}
 and the orthorhombic T$_d$ phase\cite{Clarke,Puotinen,Yan,Zandt}, the phase that the crystals of the present measurements are in at low temperatures. The orthorhombic (T$_d$) phase at low temperature is non-centrosymmetric\cite{Zhang,Clarke,Puotinen} (space group Pmn21) and this phase is achieved from the inversion-symmetric monoclinic (1T$'$) phase through a structural phase transition at $\sim$ 250 K\cite{Clarke,Brown,Zhang}. In terms of physical properties, MoTe$_2$ has recently been shown to be a type II Weyl semimetal which also superconducts below a very low critical temperature $T_c\sim$ 0.1 K in it's pristine form\cite{Wang1,Huang,Qi}. It was earlier shown that on doping the Mo-sites of MoTe$_2$ with Re, it is possible to increase the critical temperature by an order of magnitude, up to 4.1 K\cite{Mandal}. 

Apart from the enhanced superconductivity, the resistivity of polycrystalline Re$_x$Mo$_{1-x}$Te$_2$ displayed unusual normal state "hump"-structure forming around 250 K with a hysteresis in the heating vs. cooling data. While the hysteretic nature could be attributed to the structural phase transition, the hump-structure remained unexplained. Since the dichalcogenides often show a charge density wave (CDW) state\cite{Wilson,Manzeli,Shen,Kikuchi} and MoTe$_2$ also undergoes CDW transition in thin film forms\cite{Dong}, suggestions were made to attribute the hump structure with a CDW phase in Re$_x$Mo$_{1-x}$Te$_2$. The primary motivation of the present work was to probe the CDW phase, if it exists, microscopically in high quality single crystals of Re$_x$Mo$_{1-x}$Te$_2$. We have performed scanning tunneling microscopy and spectroscopy in such crystals at various temperatures. The key results are (i) observation of CDW-like phase and (ii) ferroelectric-like hysteretic phase transition in non-centrosymmetric orthorhombic Re$_x$Mo$_{1-x}$Te$_2$ (schematic of the crystal structure is shown in Figure 1(a)).

\section*{\hspace*{-6cm}{Methods}}
High quality single crystals of Re$_x$Mo$_{1-x}$Te$_2$ were used for all the measurements presented here. Single crystals of 1T-Mo$_{1-x}$Re$_x$Te$_2$ were grown by the Chemical Vapor Transport (CVT) method using iodine as the transporting agent. In the first step, pure phase polycrystalline samples of composition Re$_x$Mo$_{1-x}$Te$_2$ ($x$ =0.2) were prepared by the standard solid-state reaction process. Stoichiometric mixtures of Mo (99.9\% pure), Re (99.99\% pure), and Te (99.99\% pure) powders were ground together, pelletized and sealed in an evacuated quartz tube. The ampoules were first heated at 1100$^{o}$C for 24 hours, followed by quenching to avoid the formation of the 2H phase. After this heat treatment, samples were reground and the same heat treatment was repeated. The obtained pure polycrystalline samples were used in the crystal growth process. Crystallization was carried out from 1100$^{o}$C (hot zone) to 900$^{o}$C (cold zone) for seven days in a two-zone furnace where the polycrystalline sample was kept in the hot zone. Finally, the sealed quartz tube was quenched to avoid the formation of the hexagonal phase. The quality of the crystals were checked by performing low-energy electron diffraction (LEED) measurements on the cleaved surface of a single crystal in ultra high vacuum (UHV). Sharp spots corresponding to well defined crystal planes are clearly observed in the LEED Patterns (Figure 1(b)). 

\begin{figure}[h!]
\includegraphics[width=.4\textwidth]{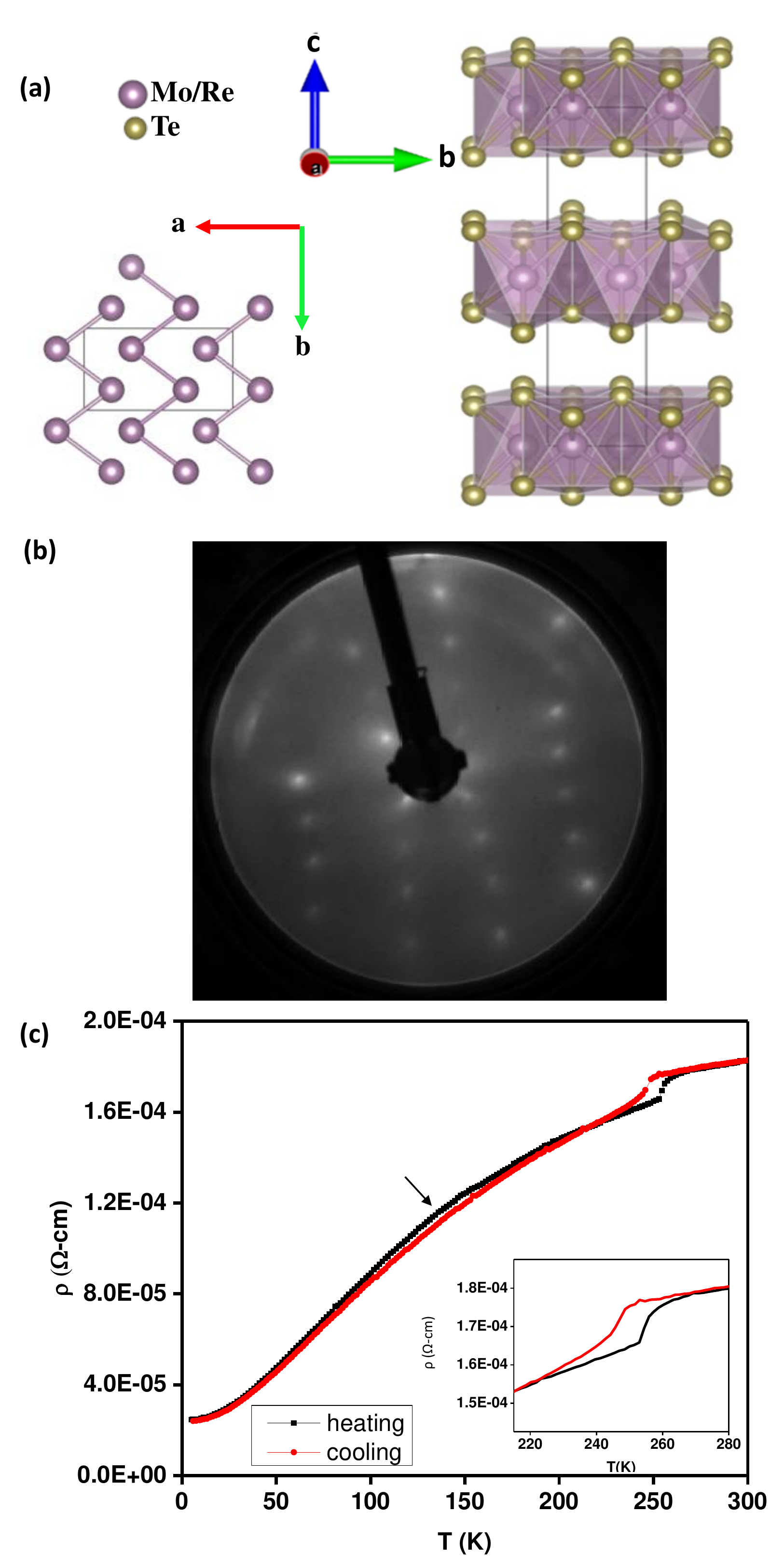}
\caption{(a) The crystal structure of Re$_x$Mo$_{1-x}$Te$_2$ in orthorhombic T$_d$ phase. (b) a LEED pattern confirming high quality of the crystal used for the measurements. (c) Resistivity ($\rho$) vs temperature (T) measured by 4-probe method on Re$_x$Mo$_{1-x}$Te$_2$.}	
\end{figure}

\begin{figure}
		\includegraphics[width=.5\textwidth]{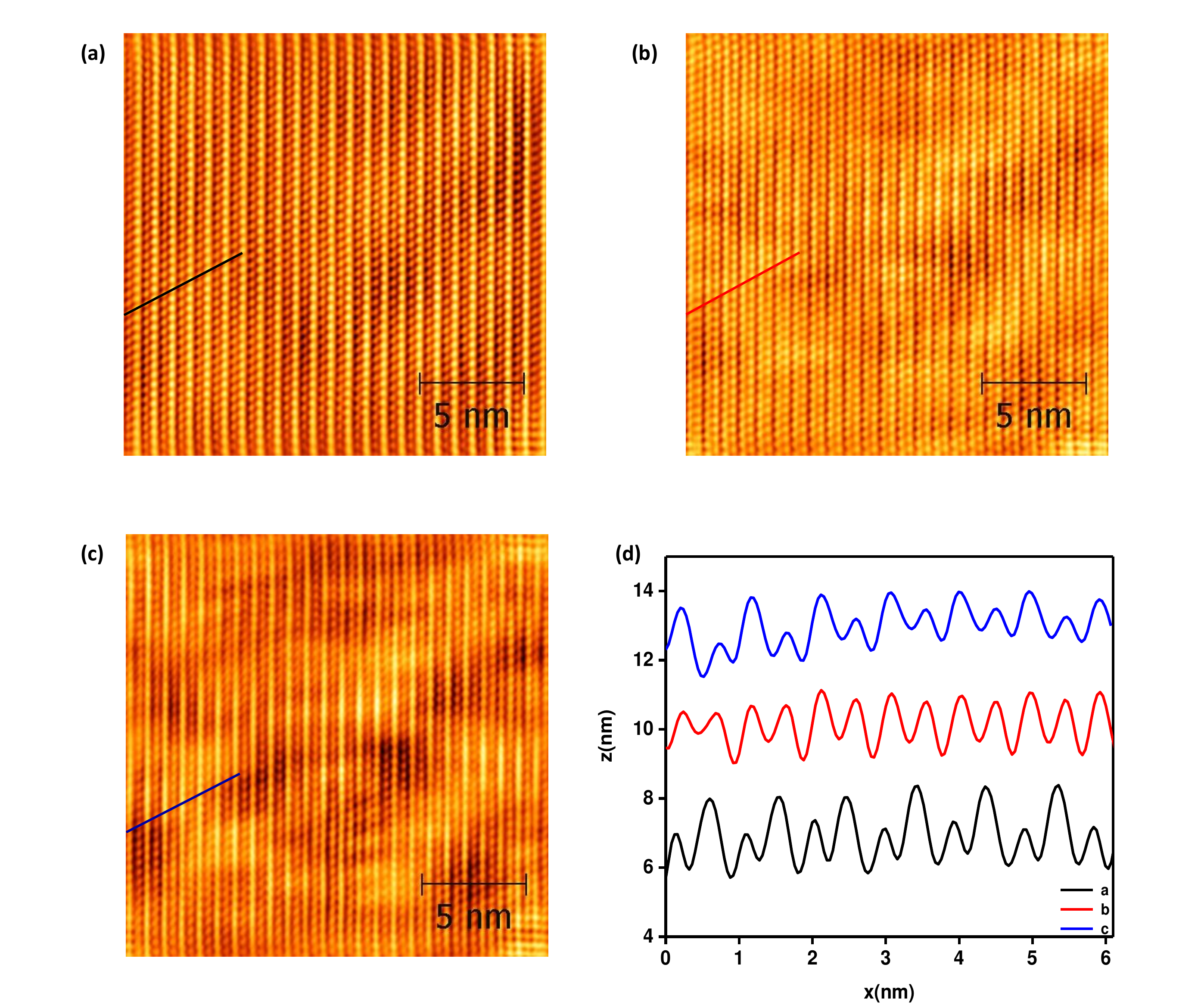}
		\caption{ Atomic resolution images for (a) $V$ = 2mV, (b) $V$ = 6 mV and (c) $V$ = 20 mV. (d) height profile corresponding to the line-cuts shown in (a,b,c).}	

\end{figure}

\section*{\hspace*{-3cm}{Results and Discussion}}

The temperature dependent electrical transport measurements were carried out in a Quantum design (QD) Physical Properties Measurement System (PPMS) employing a conventional four-probe ac technique with an excitation current of 10 mA at 17 Hz. Each temperature scan (heating and cooling) was done extremely slowly (over 12 hours) in order to avoid any artificial difference between the heating and cooling runs. The transport data are presented in Figure 1(c). The black line shows the data recorded during heating from 5 K to 300 K and the red line shows the data recorded during cooling. The cooling curve falls on the heating curve over most of the temperature range. Near 250 K, a first order jump in the $\rho$-$T$ curve is seen and near the jump, the cooling curve deviates from the heating curve. The deviation is more clearly visible in the plot shown in the $inset$ of Figure 1(c). This is a characteristic of a structural phase transition; in this case this is a transition from the 1T$^{'}$ phase to the T$_d$ phase. This was earlier observed in undoped MoTe$_2$\cite{Clarke} and polycrystalline Re$_x$Mo$_{1-x}$Te$_2$\cite{Mandal}, but in our measurements on single crystals, the effect of the structural transition is seen to be far more pronounced. Below 250 K, a hump structure is seen in the $\rho$-$T$ curve. As discussed before, such hump structure can be a signature of a CDW phase in the system\cite{Piatti,Naito,Cao}. Hence, it is important to investigate the possibility of a CDW phase in Re$_x$Mo$_{1-x}$Te$_2$ by scanning tunnelling microscopy (STM) and scanning tunnelling spectroscopy (STS).

\begin{figure}[h!]
		\includegraphics[width=.5\textwidth]{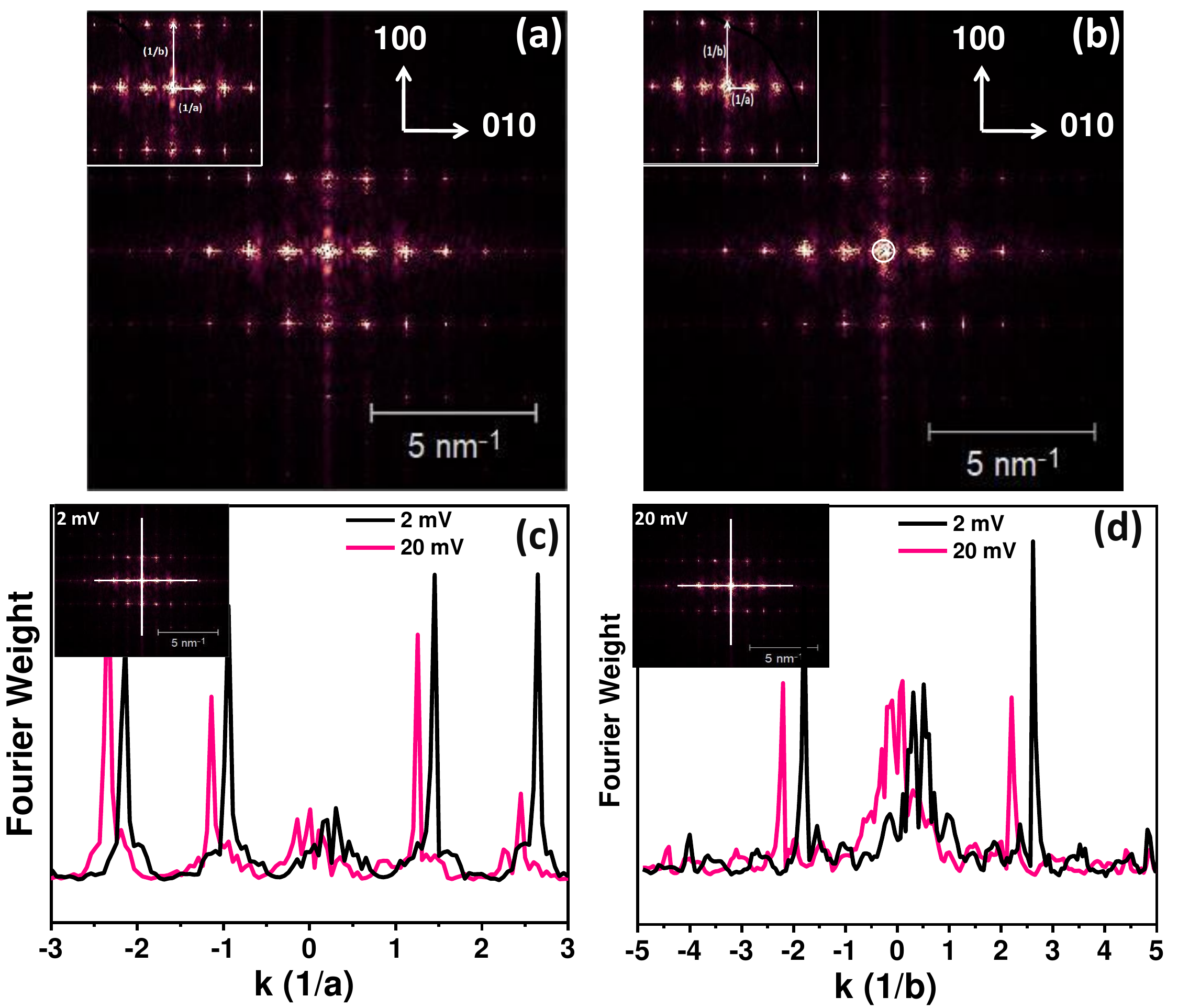}
		\caption{ FFT image (a) corresponding to the image in Figure 2(a), (b) corresponding to the image in Figure 2(c) and circle shows the site of zero - point (inset in (a) and (b) shows the corresponding reciprocal lattice vectors). (c) Intensity profiles corresponding to the lines along (010) in (a,b). (d) Intensity profiles corresponding to the lines along (100) in (a,b).}	

\end{figure}

\begin{figure}[h!]
		\includegraphics[width=.5\textwidth]{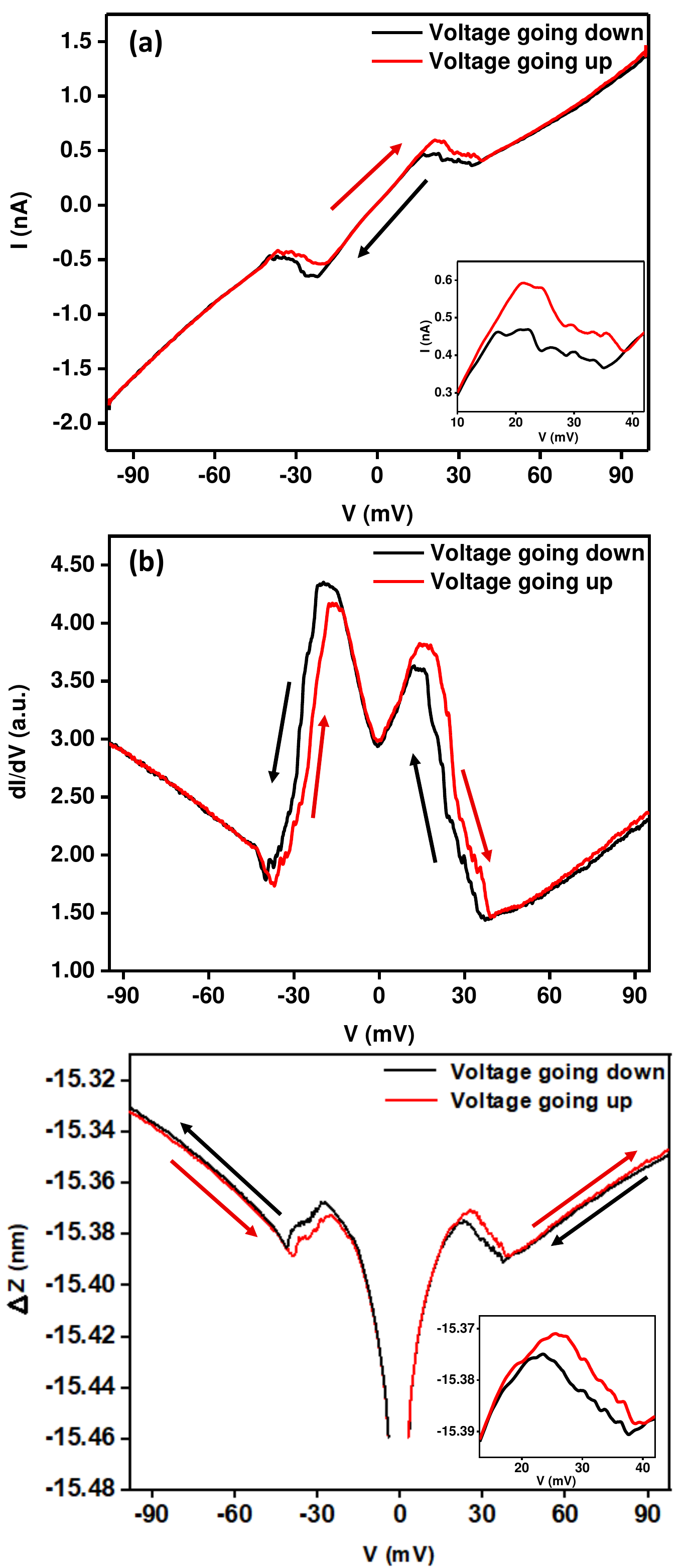}
		\caption{(a) Tunneling $I-V$. $inset$: the region showing hysteresis. (b) tunneling $dI/dV$ vs $V$ showing hysteresis. (c) $z-V$ data. $inset$: The region showing hysteresis is highlighted.}	

\end{figure}

The STM/S experiments were performed in a low temperature, ultra-high vacuum (UHV) He$^3$ cryostat working down to 300 mK (Unisoku system). Single crystals of Re$_x$Mo$_{1-x}$Te$_2$ were first mounted on a low-temperature cleaving stage housed in an exchange chamber integrated with the main STM chamber and separated by a UHV gate valve. The crystals were cleaved in the exchange chamber by an $in-situ$ cleaver while the crystal temperature was controlled at 80 K. Immediately after cleaving, the crystals were transferred by an UHV manipulator to the scanning stage which hangs from the end of the He$^3$ insert with three metal springs at sub-Kelvin temperatures. For STM measurements, the sample stage was biased and the tip was virtually grounded through the current pre-amplifier. A lock-in based modulation technique was employed to perform STS.

Atomic resolution images of the surface of Re-MoTe$_2$ captured at 5 K (in the normal state, above superconducting $T_c$) are shown in Figure 2 (a,b,c). The rectangular surface lattice is clearly seen with unidirectional chains of atoms corresponding to the chains of Te atoms along (100) as in the $T_d$ phase of MoTe$_2$. As we will discuss later, the tunnelling spectroscopy measurements showed switching behaviour with hysteresis (Figure 4), it was not possible to directly measure the CDW gap by STS. Hence, it was important to analyze the atomic resolution images carefully to extract microscopic information about the possible CDW phase here. In absence of information on the value of the CDW gap, we relied on the temperature scale of the onset of CDW-like hump structure in the transport data and estimated the gap to be $\sim$ 12 meV (this estimate was using $\Delta$/(k$_B$$T_c$) = 1. Considering the mean field ratio of  $\Delta$/(k$_B$$T_c$) = 1.7, the gap is found to be 18 meV with T$_c$ $\sim$ 130 K .)  With this information in hand, we investigated the atomic resolution images at different tip-sample bias ($V$) ranging from 2 mV to 20 mV. We show three representative images here for $V$ = 2 mV (Figure 2(a)), $V$ = 6 mV (Figure 2(b)) and $V$ = 20 mV (Figure 2(c)). The energy dependent evolution of the underlying defect states are clearly seen in the background of these images.  In terms of the relative intensity of the planes of atoms, an energy dependent shift is observed as it is shown in the line profiles in Figure 2(d) corresponding to the lines shown in the respective images.  

In order to probe any possible change in the lattice structure below and above the estimated CDW gap, we have performed fast Fourier transform (FFT) of the real space images at $V$ = 20 mV (Figure 3(a)) and at $V$ = 2 mV (Figure 3(b)). We have plotted the intensity profile along (100) (Figure 3(c)) and along (010) (Figure 3(d)) corresponding to the lines shown in Figure 3(a) and 3(b) respectively. A slight shift in the corresponding k vectors are seen for both (010) and (100) which might be due to CDW. The shift is almost same for all the spots within the lines. This and the fact that not superstructure related spot is seen in the FFT images, it can be concluded that the CDW phase in this case is dynamic in nature. 

Now we focus on the scanning tunnelling spectroscopy (STS) experiments. In Figure 4(a) we show the tunnelling $I-V$ characteristics. As $V$ is continuously varied from -90 mV to +90 mV, $I-V$ shows nonlinearity around $\pm$ 30 mV. As $V$ is decreased, suprisingly, a hysteresis appears near the non-linear portion of the $I-V$ curve. For clarity, we have shown a zoomed view of the part showing hysteresis in the $inset$. The corresponding $dI/dV$ shows two peaks symmetric in terms of the location in the $V$-scale, but asymmetric in terms of relative amplitude (Figure 4(b)). $dI/dV$ shows a negative slope with $V$ and a prominent hysteresis between 15 mV to 40 mV, as the direction of $V$ is reversed. Beyond $\pm$ 40 mV, the $dI/dV$ shows a upward turn and does not show hysteresis. 

In order to understand the origin of the hysteresis, we first need to consider the basic working principle of STM. The tunnelling current in STM can be written as $I \sim e^{(-2/\hbar)\sqrt{2m\phi}z}$\cite{Fowler, Frenkel}, where $m$ is the electron mass, $\phi$ is the work function of the material under investigation and $z$ is the tip-sample distance. In another form, $I \sim c(V/z)e^{-2\kappa z}$, where $c$ and $\kappa$ are constants. In the constant current mode of STM imaging, a computer controlled feedback loop controls $z$ in order to keep $I$ at it's set point with a bias $V$ between the tip and the sample. Now, if the surface over which the tip stands undergoes physical deformation upon application of an electric field between the tip and the sample, the tunnelling current is also expected to change as that causes a change in $z$. Such a field induced deformation ($\Delta z$) is seen in piezo-electric/ferroelectric (all ferroelctrics are also piezoelectrics) materials, where the hysteresis with electric field is naturally expected. Hence, since the crystal structure of Re$_x$Mo$_{1-x}$Te$_2$ is non-centrosymmetric, it is rational to attribute the observed hysteresis effect to local field-induced ferroelectric like switching\cite{Dragan}. 

In order to verify if a local deformation under the STM actually takes place or not, we have carried out direct measurements of $z$ vs $V$. As seen in Figure 4(c), the effect of local deformation is clearly seen. First, as $V$ increases, $z$ also increases with +ve $dx/dV$. Beyond $\sim$ 30 mV, $dz/dV$ becomes -ve and remains so upto 40 mV. Beyond this point, $dz/dV$ again becomes positive. Within the negative $dz/dV$ region, the $z-V$ curve shows hysteresis. We have shown the hysteretic part in the $inset$ for clarity. Therefore, it can be concluded with confidence that the surface of the crystal deforms locally upon the application of the STM bias and for a certain range of bias, the deformation is hysteretic and the same hysteresis is observed in the $I-V$ and $dI/dV -V$ characteristics. This matches remarkably well with the ``butterfly loop"-like behaviour that is expected on a piezo-electric/ferroelectric crystal\cite{Dragan}.\\
We note that a ferroelectric phase was earlier observed in the thin films of WTe$_2$ and T$_d$-type MoTe$_2$. However, in our case, since the bulk crystalline phase shows ferroelctric switching, it is rational to summarize that the ferroelectric-like behaviour in Re$_x$Mo$_{1-x}$Te$_2$ has a different origin which be related to the doping of Re atoms in MoTe$_2$.\\
It should also be noted that usually ferroelectric behaviour is seen in insulators. It is surprising that a ferroelectric like switching is being observed here in a seemingly good metal which is also a superconductor at lower temperatures. In the past so-called ''metallic ferroelectricity" was seen in systems like the pyrochlore Cd$_2$Re$_2$O$_7$ and LiOsO$_3$\cite{Keppens,Mandrus}. However, in the present case, the microscopic mechanism that gives rise to the ferroelectric like hysteretic behaviour in Re$_x$Mo$_{1-x}$Te$_2$ may be related to the low temperature non-centrosymmetric crystal structure. More theoretical investigations are required to gain complete understanding of the observed phenomenon.

\section*{\hspace*{-6cm}{Conclusion}}

In conclusion, from transport measurements and atomic resolution imaging and the analysis of the same, we have shown that a possible CDW phase appears in Re$_x$Mo$_{1-x}$Te$_2$. Detailed STS measurements and $z-V$ measurements reveal a local ferroelectric like switching behaviour in the crystal. The observations elucidate how further exotic physics can emerge in topological chalcogenides upon doping and thus new promises for studying competing orders in topological materials open up. The results are expected to spur further research on microscopic understanding of the phenomena presented here.

\section*{\hspace*{-4cm}{Acknowledgement}}

AV thanks UGC for senior research fellowship (SRF). GS acknowledges DST, Govt. of India for financial support through Swarnajayanti fellowship (grant number: \textbf{DST/SJF/PSA-01/2015-16}).


\begin{thebibliography}{100}

\section*{\hspace*{-5cm}{References}}
\bibitem{Motohiko}M. Ezawa, \textit{Scientific Reports}\textbf{ 9}, 5286 (2019).
 

\bibitem{Xiaoyin}Xiaoyin Li, Shunhong Zhang, Qian Wang, \textit{Nanoscale} \textbf{9}, 562 (2017).

\bibitem{Yandong}Y. Ma, L. Kou, Xiao Li, Y. Dai, S. C. Smith, and T. Heine, \textit{Physical Review B}  \textbf{92}, 085427 (2015).

\bibitem{Noh} Han-Jin Noh, J. Jeong, En-Jin Cho, K. Kim, B.I. Min and Byeong-Gyu Park, \textit{Physical Review Letters} \textbf{119}, 016401(2017).


\bibitem{Amit} Amit, R.K. Singh, Neha Wadehra, S.  Chakraverty, Yogesh Singh,
\textit{Physical Review Materials} \textbf{2}, 114202 (2018).

\bibitem{Shekhar} S. Das, Amit, A. Sirohi, L. Yadav, S. Gayen, Y. Singh, and G. Sheet, \textit{Physical Review B}\textbf{ 97}, 014523 (2018).

\bibitem {Qi}Y. Qi et al., \textit{Nature Communications} \textbf{7} , 11038 (2016).

\bibitem{Liu} Junwei Liu, Hua Wang, Chen Fang, Liang Fu, and Xiaofeng Qian,
\textit{Nano Letters}, \textbf{17}, 1, 467-475  (2017).

 \bibitem{Revolinsky} E. Revolinsky, and D. Beerntsen, \textit{Journal of Applied Physics} \textbf{35}, 2086 (1964).

\bibitem{Deng}K. Deng, G. Wan, P. Deng, K. Zhang, S. Ding, E. Wang, M. Yan, H. Huang, H. Zhang, Z. Xu, J. Denlinger, A. Fedorov,H. Yang, W. Duan, H. Yao, Y. Wu, S. Fan, H. Zhang, X. Chen, and S. Zhou,  \textit{Nature Physics} \textbf{12}(12), 1105- 1110 (2016).

\bibitem{Tamai}A. Tamai, Q. S. Wu, I. Cucchi, F. Y. Bruno, S. Ricco, T. K. Kim, M. Hoesch, C. Barreteau, E. Giannini, C. Besnard, A. A. Soluyanov, F. Baumberger,\textit{ Physical Review X} \textbf{6}(3), 031021 (2016).

\bibitem{Jiang} J. Jiang, Z. K. Liu, Y. Sun, H. F. Yang, C. R. Rajamathi, Y. P. Qi, L. X. Yang, C. Chen, H. Peng, C. C. Hwang, S. Z. Sun, S. K. Mo, I. Vobornik, J. Fujii, S. S. P. Parkin, C. Felser, B. H. Yan, and Y. L. Chen,\textit{ Nature Communications}\textbf{ 8}, 13973 (2017).

\bibitem{Yan} Y. Sun, Shu-Chun Wu, Mazhar N. Ali, Claudia Felser, and Binghai Yan. \textit{Physical Review B} \textbf{92}, 161107(R) (2015).

\bibitem{Guguchia} Z. Guguchia et. al. \textit{Nature Communications} \textbf{8}, 1082 (2017).

\bibitem{Puotinen} D. Puotinen and R. E. Newnham, \textit{Acta Crystallography} \textbf{14}, 691 (1961).

\bibitem{Albert} H. Albert, R. Kershaw, K. Dwight and A. Wold,, \textit{Solid State Communications} \textbf{81}, 649- 651 (1992).

\bibitem{Vellinga} M. B. Vellinga, R. De Jonge,  C. Haas,  \textit{Solid State Chemistry} \textbf{2}, 299-302 (1970).

\bibitem{Dawson}W. G. Dawson and D. W. Bullett \textit{Journal of Physics C: Solid State Physics} \textbf {20}, 6159-6174  (1987).

\bibitem{Brown} B. E. Brown.\textit{ Acta Crystallography} \textbf{20}, 268-274 (1966).

\bibitem{Clarke} R. Clarke, E. Marseglia, and H. P. Hughes,\textit{ Philosophical
Magazine B} \textbf{38}, 121-126 (1978).

\bibitem{Zandt} T. Zandt, H. Dwelk, C. Janowitz, and R. J. Manzke,\textit{ Journal of Alloys \& Compound} \textbf {442}, 216 (2007).

\bibitem{Zhang} K. Zhang, C. Bao, Q. Gu, X. Ren, H. Zhang, K. Deng, Y. Wu, Y. Li, Ji Feng , S. Zhou, \textit{Nature Communications} \textbf{7}, 13552 (2016)

\bibitem {Wang1}Z. Wang, D. Gresch, A. A. Soluyanov, W. Xie, S. Kushwaha, and X. Dai, \textit{Physical Review Letters} \textbf{117}, 056805 (2016).



\bibitem{Huang} L. Huang, T. M. McCormick, M. Ochi, Z. Zhao, M.-To Suzuki,R. Arita, Y. Wu, D. Mou, H. Cao, J. Yan, N. Trivedi and A. Kaminski, \textit{Nature Materials} \textbf{15}, 1155-1160 (2016).


\bibitem{Mandal}M. Mandal, S. Marik, K. P. Sajilesh, Arushi, D. Singh,
J. Chakraborty, N. Ganguli, and R. P. Singh,\textit{Physical Review
Materials} \textbf{2}, 094201 (2018).

\bibitem{Wilson} J. A. Wilson, F. J. Di Salvo, and S. Mahajan,
\textit{Physical Review Letters}\textbf{ 32}, 882 (1974).

\bibitem{Manzeli}S. Manzeli, D. Ovchinnikov, D. Pasquier, O. V. Yazyev , A. Kis, \textit{Nature Reviews Materials} \textbf{2}, 17033 (2017).

\bibitem{Shen} D. W. Shen, B. P. Xie, J. F. Zhao, L. X. Yang, L. Fang, J. Shi, R. H. He, D. H. Lu, H. H. Wen, and D. L. Feng, \textit{Physical Review Letters} \textbf{99}, 216404 (2007)


\bibitem{Kikuchi}A. Kikuchi , M. Tsukada, \textit{Surface Science} \textbf {326}, 195-207 (1995).

\bibitem{Dong} L. Dong et. al.\textit{ Chinese Physics Letters} \textbf{6}, 35, 066801  (2018).

 
\bibitem{Piatti} E. Piatti, Q. Chen, M. Tortello, J. Ye, and R. S. Gonnelli,\textit{ Applied
Surface Science} \textbf{461} , 269-275 (2018).

\bibitem{Naito} M. Naito, and S. Tanaka, \textit{Journal of the Physical Society of Japan} \textbf{51}, 219-227 (1982).

\bibitem{Cao} Zi-Yu Cao, Jia-Wei Hu, A. F. Goncharov, and Xiao-Jia Chen,
\textit{Physical Review B} \textbf{97}, 214519 (2018).

\bibitem{Fowler}R. H. Fowler and L. Nordheim,\textit{ Proceedings of the  Royal Society of London A}, \textbf{119}, 173 (1928)

\bibitem{Frenkel} J. Frenkel, \textit{Physical Review} \textbf{36}, 1604 (1930).

\bibitem{Dragan} Dragan Damjanovic, I. Mayergoyz and G.Bertotti (Eds.); \textit{The Science of Hysteresis}, Volume \textbf{3}, Elsevier (2005).

\bibitem{Keppens} Veerle Keppens, \textit{Nature Materials} \textbf{12}, 952–953 (2013).

\bibitem{Mandrus}I. A. Sergienko, V. Keppens, M. McGuire, R. Jin, J. He, S. H. Curnoe, B. C. Sales, P. Blaha, D. J. Singh, K. Schwarz, and D. Mandrus, \textit{Physical Review Letters} \textbf{92 (6)}, 065501 (2004).

\end{thebibliography}
\end{document}